\documentclass[usegraphicx]{mn2e}
\def\H0{{\it H}$_0$}
\def\Ms{{\it M}$_\odot$}

\def\q0{{\it q}$_0$}
\def\kmps{km~s$^{-1}$}

\def\htwoo{H$_2$O}
\def\ergps{erg~s$^{-1}$}

\def\Ms{{\it M}$_\odot$}

\def\deg{$^{\circ}$}

\def\nH{$N_{\rm H}$\thinspace} 

\def\psqcm{cm$^{-2}$}
\def\pyr{yr$^{-1}$}
\def\ergpspsqcm{erg~cm$^{-2}$~s$^{-1}$}

\def\phpspsqcm{ph\thinspace s$^{-1}$\thinspace cm$^{-2}$}

\def\etal{et al.\ }

\title[A Chandra observation of IC2560] 
{A Chandra observation of the H$_2$O megamaser IC 2560} 
\author[Iwasawa, Maloney \& Fabian] 
{\parbox[]{6.5in} {K. Iwasawa$^1$\thanks{E-mail: ki@ast.cam.ac.uk},
    P.R. Maloney$^2$ and A.C. Fabian$^1$}\\ 
\\
$^1$Institute of Astronomy, Madingley Road, Cambridge CB3 0HA\\ 
$^2$CASA, University of Colorado, Boulder, CO 80309-0389 USA\\ 
}
\date{}

\begin{document}

\maketitle

\begin{abstract}
  A short Chandra ACIS-S observation of the Seyfert 2 galaxy IC 2560,
  which hosts a luminous nuclear water megamaser, shows: 1) the
  X-ray emission is extended; 2) the X-ray spectrum shows emission
  features in the soft ($E < 2$ keV) X-ray band; this is the major
  component of the extended emission; and 3) a very strong (EW$ \sim
  3.6$ keV) iron K$\alpha$ line at 6.4 keV on a flat continuum. This
  last feature clearly indicates that the X-ray source is hidden
  behind Compton-thick obscuration, so that the intrinsic hard X-ray
  luminosity must be much higher than the observed, probably close to
  $\sim 3\times 10^{42}$\ergps. We briefly discuss the implications
  for powering of the maser emission and the central source.
\end{abstract}


\section{introduction}

IC 2560 is a relatively nearby ($D=26$ Mpc) barred spiral galaxy,
classified as a Seyfert 2 (Fairall 1986; Kewley et al 2001). It is notable for
exhibiting luminous \htwoo\ maser emission from its nucleus (Braatz,
Wilson, \& Henkel 1996). This maser emission resembles that from the
archetypal water megamaser source NGC 4258 in that high-velocity
emission is seen up to $\Delta V\approx 400$ \kmps\ away from the
systemic velocity, the systemic emission is much stronger than the
high-velocity emission, and centripetal acceleration of systemic
velocity features has been reported (Ishihara \etal 2001). The
high-velocity emission has not yet been imaged; if the data are
interpreted in the framework of a Keplerian disk, as in NGC 4258, the
implied central mass is $M_c\approx 2.8\times 10^6$ \Ms. Ishihara
\etal also analyzed an ASCA observation of IC 2560 and concluded that
it possesses a fairly heavily obscured ($N_H\sim 3\times 10^{23}$
\psqcm) but low luminosity ($L_{2-10 \rm keV}\sim 10^{41}$ \ergps)
X-ray source.

In this paper we report on a short Chandra observation of IC 2560,
which reveals spatially extended soft X-ray emission and a
substantially different nature to the hard X-ray source than was
inferred by Ishihara \etal (2001) from the ASCA data. Throughout we
assume the distance to the galaxy to be 26 Mpc, giving an angular
scale of 120 pc arcsec$^{-1}$.

\section{Observation and data reduction}

IC 2560 was observed with Chandra X-ray Observatory (hereafter
Chandra, Weisskopf et al 2000) on 2000 October 29-30, using the
Advanced CCD Imaging Spectrometer (ACIS). The galaxy was positioned on
the back-illuminated CCD, ACIS-S3 detector. The focal plane
temperature was $-120^{\circ}$ during this observation. The data
reduction was carried out using the Chandra Interactive Analysis of
Observation (CIAO) version 2.2 package and calibration files in the
calibration databese (CALDB) version 2.10. The S3 detector has been
inspected for background flares using a source-free region in the
2.5--7 keV band. The detector background appears to be stable during
this observation: the background light curve shows that all the data
points are within 30 per cent of the mean value. Since the X-ray
source of interest is barely extended beyond the point spread function
(see Section 3.1), little impact from such moderate background flares
is expected on the spectral data of the source.  Therefore no
background flare rejection has been applied and the resulting good
exposure time is 9.8 ks. The mean count rate of the source, corrected
for the background, is $3.2\times 10^{-2}$ ct s$^{-1}$, sufficiently
low that the data are unaffected by pile-up. An aspect offset, which
was present in the original event file, has been corrected using the
latest 2002-May-02 alignment file; the observed X-ray emission peaks
at the position R.A.= 10$^{\rm h}$16$^{\rm m}$18.7$^{\rm s}$, Dec.=
--33\deg 33\arcmin 49\farcs 6s (J2000), about 0.2 arcsec away from the
nuclear position in NED\footnote{NASA/IPAC Extragalactic Database
  (NED) which is operated by the Jet Propulsion Laboratory, California
  Institute of Technology, under contract with the National
  Aeronautics and Space Administration.} 
and well within the absolute
astrometric uncertainty ($\sim 0.6$ arcsec) of Chandra pointings.  The
spectral analysis presented below was performed with the spectral
analysis software, XSPEC version 11.2.

\section{results}

\subsection{Extended X-ray emission}

%Fig. 1 -- full band image
\begin{figure}
\centerline{\includegraphics[width=0.4\textwidth,angle=0,
    keepaspectratio='true']{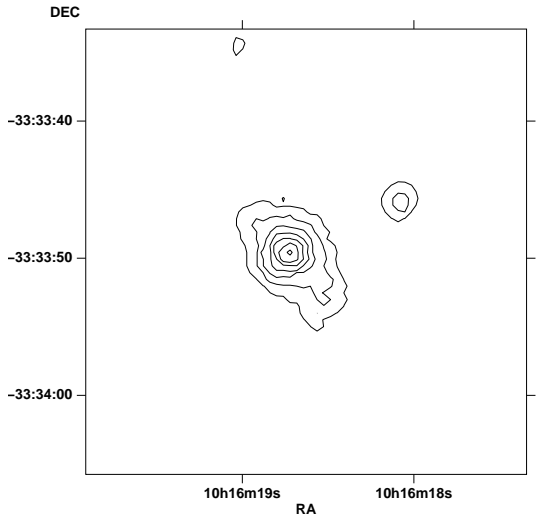}} 
\caption{
  The Chandra ACIS-S3 image of IC2560 in the 0.4--7 keV band. The
  contours are drawn at seven logarithmic intervals from 2 per cent to
  80 per cent of the X-ray peak at the nucleus position. Low surface
  brightness extension appears to be present in the NE-SW direction.
  The sky coordinates are of J2000. The angular scale is $\approx
  120$ pc arcsec$^{-1}$.}
\end{figure}

% radial profile -- Fig.2

\begin{figure}
\centerline{\includegraphics[width=0.35\textwidth,angle=270,
    keepaspectratio='true']{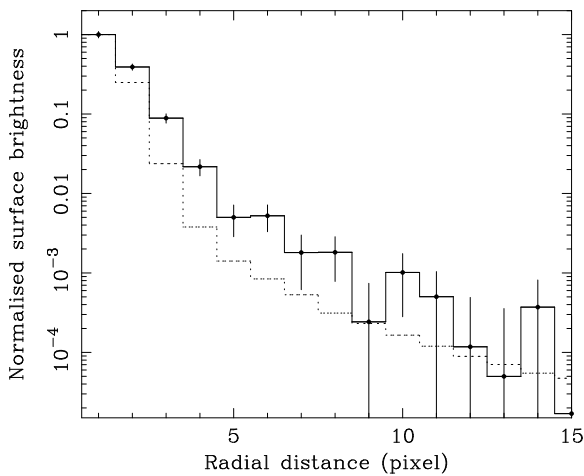}} 
\caption{
  The radial surface brightness profiles of the 0.4--7 keV emission
  from IC 2560 (filled circles with solid line histogram) and of a
  point spread function computed at 1 keV (dotted-line histogram). The
  data for IC 2560 has been corrected for backgroud and the two
  profiles are normalized by the peak brightnesses, respectively.
  One pixel corresponds to $\approx 0.5$
  arcsec or 60 pc. }
\end{figure}

X-ray emission from IC2560 is found to be extended in the Chandra
image.  Fig. 1 shows the 0.4--7 keV band ACIS-S3 image of IC2560. The
image suggests that there are a faint extension to the SW up to 6
arcsec ($\sim 720$ pc), and possibly to the NE up to 5 arcsec ($\sim
600$ pc), although the significance of these features is low ($\sim
2\sigma$). The azimuthally-averaged radial surface brightness profile
of the image in Fig. 1 is shown in Fig. 2. To investigate the
extension of the X-ray core, a point spread function (PSF) is computed
and plotted in Fig. 2 for a comparison. The PSF was computed for the
same position on the detector as that of IC2560 in this observation,
and the energy, at which the PSF was constructed, was assumed to be 1
keV, as much of the detected counts are distributed around that energy
(see the energy spectrum in Fig.  3). A simple comparison with the PSF
indicates that the core region of IC2560 is extended with a
significance of larger than $3\sigma$ and at least by 80 pc in radius.
The hard band (3--7 keV) image shows a possible elongation in the
SE--NW direction, but with only 58 counts detected in the energy
range, the reality of this feature is highly uncertain.

\subsection{Spectrum}

% Fig. 3 -- spectrum

\begin{figure}
\centerline{\includegraphics[width=0.4\textwidth,angle=270,
keepaspectratio='true']{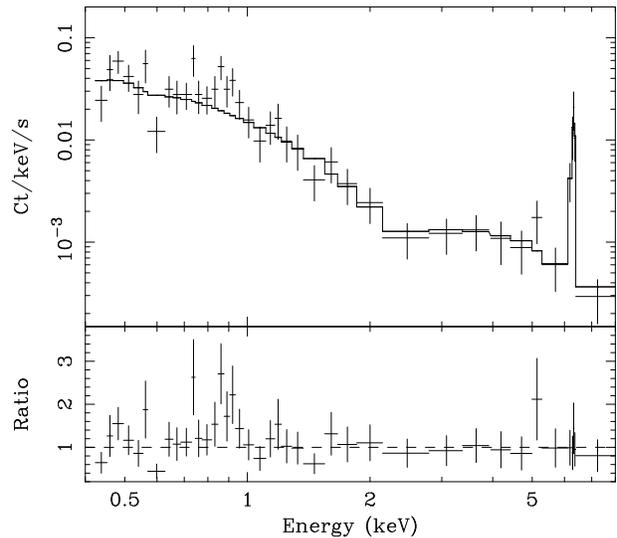}}
\caption{The ACIS-S spectrum of IC2560. The solid histogram shows the
best-fitting model consisting of a broken power-law and a gaussian
line for Fe K, modified only by Galactic absorption. Note the residuals
in the soft X-ray range (see also Fig. 4).}
\end{figure}

Fig. 3 shows the Chandra ACIS-S spectrum extracted from a circular
region with radius of 2.5 arcsec centred on the X-ray peak.  A very
strong Fe K$\alpha$ line at 6.4 keV is clearly seen on a faint, flat
continuum. A steep rise of soft X-ray emission is seen below 2 keV,
which appears to be the major component of the extended emission. When
the 0.4--7 keV data are fitted by a broken power-law modified only by
Galactic absorption\footnote{This value is obtained by averaging the
  data of the HI map over a 1 degree$^2$ field at the position of the
  galaxy, and may not be an exact column density towards the position
  of IC2560. However, any fluctuations within the field are not
  expected to be large enough to affect our results.} (\nH $ =
6.5\times 10^{20}$\psqcm, Dickey \& Lockman 1990) plus a gaussian for
the Fe K line (Fig. 3), a spectral break is found at $E_{\rm
  br}=2.0^{+0.5}_{-0.4}$ keV with photon indices of $\Gamma
=2.8^{+0.2}_{-0.2}$ and $\Gamma = 0.5^{+0.3}_{-0.7}$ below and above
the break energy, respectively (the quoted errors throughout are 90
per cent confidence limits for one parameter of interest). The line
centroid of the Fe K emission is $6.41\pm 0.03$ keV when corrected for
the galaxy's redshift. The line is not spectrally resolved: the upper
limit to the dispersion of a gaussian is 150 eV. Besides the Fe K
line, the residuals of the above fit suggest the presence of soft
X-ray emission features, the origin of which is discussed later.  The
fluxes as observed are $5.7\times 10^{-14}$ \ergpspsqcm\ in the 0.5--2
keV band, and $3.6\times 10^{-13}$ \ergpspsqcm\ in the 2--10 keV band.
The 0.5--2 keV luminosity corrected for Galactic absorption is
$5.0\times 10^{39}$ \ergps. The observed 2--10 keV luminosity is
$2.7\times 10^{40}$ \ergps, of which half is contributed by the iron K
line.

The flat hard continuum is consistent with a cold reflection spectrum
(e.g., Iwasawa, Fabian \& Matt 1997 for an example seen in NGC1068).  A
spectrum of reflection alone from an optically thick cold slab (e.g.,
computed using {\tt pexrav} by Magdziarz \& Zdziarski 1995 in XSPEC,
assuming the incident source to have a power-law spectrum with $\Gamma
= 2$) is in good agreement with the continuum shape above 3 keV. The
Fe K line flux is $(1.32\pm 0.55)\times 10^{-5}$ \phpspsqcm, and its
equivalent width with respect to the reflection continuum is extremely
large, $3.6\pm 1.5$ keV.

% soft lines -- Fig 4

\begin{figure}
\centerline{\includegraphics[width=0.4\textwidth,angle=270,
keepaspectratio='true']{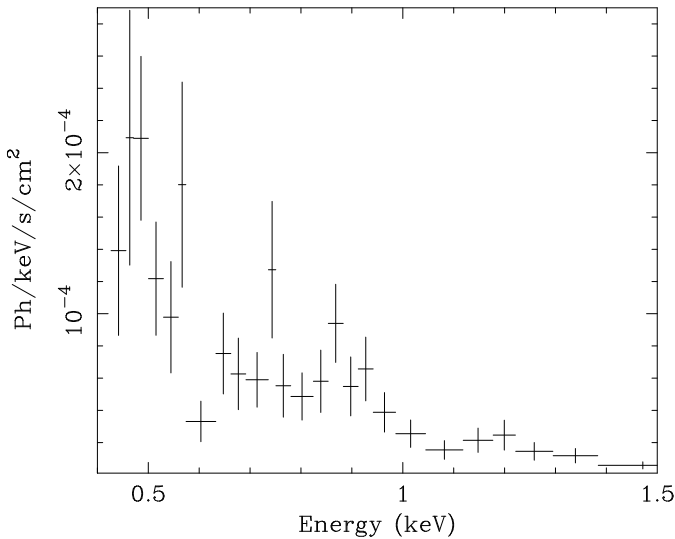}}
\caption{The low energy part of the spectrum of IC2560. The data have
been corrected for the detector efficiency and the energy scale has
been corrected for the galaxy's redshift ($z=0.00975$).}
\end{figure}

Based on the possible emission line features, the soft excess emission
could be thermal emission (i.e., collisionally ionized plasma) arising
from a nuclear starburst. A fit with a thermal emission spectrum
(computed by {\tt mekal} in XSPEC based on the model calculations of
R. Mewe and J. Kaastra with updated Fe L calculations by D.A. Liedahl;
Mewe Greonenschild \& van den Oord 1985; Kaastra 1992; Liedahl,
Osterheld \& Goldstein 1995) gives a temperature of $kT =
0.63^{+0.11}_{-0.11}$ keV and $0.03^{+0.03}_{-0.02}$ solar
metallicity. Since the implied metallicity from the fit is much too
low to be realistic, this single temperature thermal emission model
may not be appropriate. It is unclear whether there is a nuclear
starburst in IC2560 capable of producing the thermal emission.
Although a number of HII regions distributed along spiral arms have
been imaged with H$\alpha $+[NII] (Tsvetanov \& Petrosian 1995), these
HII regions are located 20 arcsec or more away from the nucleus where
the X-ray emission is observed, and none of them are detected in the
Chandra image. The optical light from the nucleus is dominated by
Seyfert 2 emission. The 1.6 $\mu $m luminosity of the unresolved
nucleus, $2.3\times 10^{40}$\ergps, as measured with the Near-Infrared
Camera and Multiobject Spectrometer (NICMOS) on Hubble Space Telescope
(Quillen et al 2001), is as low as typically seen in the non-Seyfert
control sample, suggesting that the near infrared emission from the
nucleus of IC2560 is largely due to a stellar cluster. Since the
near-infrared to soft X-ray luminosity ratio is more than an order of
magnitude too low to match those of starburst galaxies, a nuclear
starburst does not appear to be powerful enough to produce the
observed soft X-rays. An alternative source for the soft X-ray
emission is extended photoionized gas. A few suggestive features in
the energy range between 0.5--1.4 keV (Fig. 4) could be due to
recombination lines and radiative recombination continua from highly
ionized N, O, Ne, Fe (L-shell emission) and Mg. The steep rise of soft
X-ray emission towards lower energies could be dominated by these and
other emission features, which are not resolved at the spectral
resolution of a CCD, and a scattered nuclear continuum may not
necessarily be present.  Higher resolution spectroscopy is required to
test this hypothesis.

We analyzed the ASCA data of IC2560, available from the public archive,
and confirmed the presence of a strong iron K line with $EW = 2.2\pm
0.8$ keV at an energy of 6.4 keV in the spectrum. Prior to our Chandra
observation, two contrary claims have been reported for the ASCA data:
Risaliti, Maiolino \& Salvati (1999) found an enormous iron line with $EW =
6.3^{+2.6}_{-3.0}$ keV at $6.56^{+0.25}_{-0.15}$ keV, suggesting a
Compton-thick source, while Ishihara et al (2001) interpret the hard
X-ray emission as an absorbed power-law with \nH $\sim 3\times
10^{23}$ \psqcm\ without noting the iron line. Our Chandra data support
the interpretation of Risaliti et al (1999), although our ASCA results do
not agree exactly.

\section{Discussion}

\subsection{The true luminosity of the hidden active nucleus}

The hard X-ray spectrum dominated by an iron K line indicates the
absence of direct continuum emission from a central source in the
Chandra band, meaning that we are seeing only reflected light from a
hidden nucleus. The absorption column density must be larger than \nH
$\sim 1\times 10^{24}$ \psqcm, i.e., the X-ray source is Compton-thick.
How much of the intrinsic luminosity emitted by the hidden nucleus is
seen in reflection depends on the obscuration/reflection geometry.
Reflection from cold material, as inferred from the iron line energy
and the flat hard X-ray continuum, is not an efficient process, since
photoelectric absorption within the reflector suppresses reflected
light significantly, in particular at low energies. With the observed
continuum luminosity of $1.3\times 10^{40}$ \ergps\ in the 2--10 keV
band, the minimum incident luminosity to yield the reflection is about
$2.6\times 10^{41}$ \ergps. However, in a realistic toroidal geometry
in which the incident source is hidden from our direct view, the true
value will be much larger.
%Therefore the true 2--10 keV luminosity may well be of $the order of
%$10^{42}$\ergps.   

We estimate the upper limit to the 2--10 keV luminosity to be $\sim
3\times 10^{42}$\ergps, 10 per cent of the infrared ($\approx$
bolometric) luminosity, $L_{\rm 8-1000\mu m}\simeq 3\times 10^{43}$
\ergps\ (obtained from the Infra-Red Astronomical Satellite (IRAS) measurements using the formula given
in Sanders \& Mirabel (1996)), assuming the typical 2-10 keV X-ray to
bolometric luminosity ratio for Seyfert galaxies (e.g., Mushotzky,
Done \& Pounds 1993). The AGN luminosity is likely to be close to the
above upper limit, given the warm IRAS colour ($S_{60\mu {\rm
    m}}/S_{25\mu {\rm m}}=3.4$) and the AGN-dominated nuclear optical
spectrum, suggesting that the infrared emission is predominantly
powered by a hidden active nucleus.

\subsection{Large EW of Fe K line}

The equivalent width for the iron K$\alpha $ line, in excess of 3 keV,
is one of the largest measured among the reflection-dominated Seyfert 2
galaxies (see Matt et al 2001 and Levenson et al 2002 for recent
compilations).  Iron overabundance (2--3 solar) could be a possible
reason (Ballantyne, Fabian \& Ross 2002), but even with solar
metallicity, an optically thick torus can produce a very large EW. In
models assuming obscuring matter in a toroidal form, the Fe K line EW
depends on the optical depth and geometry of the torus (Leahy \&
Creighton 1993; Ghisellini, Haardt \& Matt 1994; Krolik, Madau \&
\.Zycki 1994; Levenson et al 2002). All these models indicate that, in
order to produce an EW as large as 3.6 keV, a torus needs to have a
column density \nH $\geq 3\times 10^{24}$ \psqcm (or Thomson optical
depth, $\tau_{\rm T}\geq 2$) and a small half-opening angle,
$\theta\leq 20^{\circ}$, and to be viewed nearly edge-on.

The required column density is consistent with the lack of transmitted
primary source emission in the Chandra spectrum, from which the lower
limit of the line of sight absorption column density, \nH $\geq
1\times 10^{24}$\psqcm, has been derived (Section 4.1). The
requirement of a small opening angle means that the X-ray absorbing
matter should be located close to the central source. The inner radius
of the water maser disk was inferred to be 0.07 pc (Ishihara et al
2001), but the accretion disk could extend inwards, with the inner
part of the disk too cold to induce masing (Neufeld \& Maloney 1995,
hereafter NM95). Perhaps this dense ($n>10^{10}$ cm$^{-3}$) inner part
of the accretion disk may be the region where the X-ray absorption
primarily occurs. However, in this case, the water maser-emitting part
of the disk must be warped in order to see the X-ray flux directly,
otherwise the X-rays from the central source will not impinge on the
disk.

\subsection{Water maser disk and accretion flow}

If the geometry of the maser emission in IC 2560 is similar to that in
NGC 4258, which has not yet been determined by imaging, then we can
use the X-ray emission and the kinematic data to place constraints on
fueling of the central massive black hole. The observed velocities of
maser emission in IC 2560 imply that the outer disk radius is at
$R_{out}\approx 0.26$ pc. Assuming that the maser emission is powered
by the X-ray flux from the central source (Neufeld, Maloney \& Conger
1994; NM95), as appears likely for
the majority of \htwoo\ megamaser sources (see Maloney 2002 for a
recent review), we can identify the outer radius of the disk with the
location of the molecular to atomic phase transition (see equation 4
of NM95). For a hard X-ray luminosity of $L_{2-10}=10^{42}L_{42}$
\ergps\ and black hole mass $M_c=10^6 M_6$ \Ms, the ratio of the mass
accretion rate to the viscosity parameter $\alpha$ is then
\begin{equation}
{\dot M\over \alpha}\approx 6.2\times 10^{-3} \left({R_{out}\over
  0.26\;{\rm pc}}\right)^{1.23} L_{42}^{0.53}M_6^{-0.77}\
M_\odot\;{\rm yr^{-1}} 
\end{equation}
which, for the inferred mass of the central black hole $M_c\approx
2.8\times 10^6$ \Ms, is $\dot M/\alpha\approx 2.8\times 10^{-3}$
\Ms~\pyr. Assuming that the bolometric luminosity is approximately ten
times the 2--10 keV X-ray luminosity, the product of $\alpha$ and the
radiative efficiency factor $\epsilon$ is then $\alpha\epsilon\sim
0.06$, similar to the value inferred for NGC 4258 (NM95). (In making
this estimate we have assumed that flow through the disk is steady
over the accretion timescale from the disk outer edge, and that the
disk sees the true X-ray luminosity.) Although the fractional
Eddington luminosity ($\sim 0.03$) and mass accretion rate are much
higher than in NGC 4258, the efficiency of radiation appears to be
similar. We also note that the Compton-thick obscuration of the hard
X-ray source is consistent with the accretion disk itself acting as
the obscurer, as in NGC 4258, given the larger derived value of $\dot
M/\alpha$.

\section*{Acknowledgements}
ACF and KI thank Royal Society and PPARC, respectively, for support.
PRM is supported by the National Science Foundation under grant AST
99-00871 and by NASA through Chandra X-Ray Observatory grant
GO2-3112X.

\end{document}